\documentclass[preprint,12pt]{elsarticle}

\usepackage{amssymb}
\usepackage{amsmath}

\journal{Physics Letters B}

\begin{document}

\begin{frontmatter}



\title{Analytical Study for Primordial Non-Gaussianity in the gravity 4D Einstein-scalar-Gauss-Bonnet Inflation}


\author[label1]{Afiq Agung} 

\affiliation[label1]{organization={Physics, Universitas Sulawesi Barat},
            addressline={Jl. Prof. Dr. Baharuddin Lopa, S.H, Talumung, Baurung, Kec. Banggae Timur}, 
            city={Majene},
            postcode={91412}, 
            country={Indonesia}}

\author[label2]{Usman Sambiri} 

\affiliation[label2]{organization={Physics, Universitas Khairun},
            addressline={Jl. Pertamina, Kampus II Unkhair Gambesi, Kec. Ternate Selatan}, 
            city={Ternate},
            postcode={97719}, 
            country={Indonesia}}

\author[label3]{Getbogi Hikmawan} 

\author[label3]{Freddy P. Zen} 

\affiliation[label3]{organization={Theoretical High Energy Physics Group, Department of Physics, Institut Teknologi Bandung},
            addressline={Jl. Ganesa 10, Kec. Coblong}, 
            city={Bandung},
            postcode={40132}, 
            country={Indonesia}}

\begin{abstract}
An inflationary model can be constrained by non-gaussian statistics as a parameter in the LSS (Large Scale Structure) distribution, and in the radiation of CMB (Cosmic Microwave Background) fluctuating temperature. Data on the CMB from Planck Collaboration provide up-to-date constraints on the parameters controlling the degree of non-Gaussianity in certain inflationary models, thus supporting or not supporting the model. Setting the non-Gaussianity parameter investigated in this study can be a reference whether or not it is a good parameter in constraining cosmological inflation models. This study attempts to examine the non-Gaussianity of the 3+1-dimensional 4DEGB gravitational cosmological inflation model starting from random field statistics. The non-Gaussian signature generated by the model is quantified, and the parameters controlling the degree of non-Gaussianity are constrained using data observation of Planck Collaboration. The method used in investigating non-Gaussianity is the in-in formalism, applied after obtaining the 3-point of $\zeta$ (curvature perturbation) terms of the perturbation expansion to the third order. The 3-point correlation function helps to create a bispectrum used to investigate the non-gaussinity of the inflation model. The results of this study show that the model tested is the slow roll pressed in the squeezed limit, because it witnesses a dominant local shape function. It has such as the non-gaussianity possessed by the single scalar field inflation as confirmation that Gauss-Bonnet term within Einstein-Hilbert action is topologically invariant in $D<5$ spacetimes.
\end{abstract}


\begin{highlights}
\item To study analytically how Primordial Non-Gaussianity imposes constraints on the 4D Einstein-scalar-Gauss-Bonnet gravity model.
\item This non-gaussianity was obtained from higher-order interactions, using the ‘in-in’ (the Schwinger-Keldysh) perturbative formalism, where the two-point function serves as the foundational input.
\item The dominant shape function in the result of analysis is $S^{local}$, and the appropriate limit for evaluating this function is $k_1\ll k_2 \approx k_3$.
\item This study finds that generating non-Gaussianity in the model is slow-roll parameter suppression in the squeezed limit—a behavior analogous to that of single-field scalar inflation, as confirmation that Gauss-Bonnet term within Einstein-Hilbert action is topologically invariant, and does not influence gravitational field equations in $D<5$ spacetimes.
\end{highlights}


\end{frontmatter}



\section{Introduction}
\label{sec1}

Developing a consistent theoretical framework for the inflationary epoch remains a central challenge in modern cosmology. Broadly, two main approaches have been proposed to characterize this period: the single-scalar-field paradigm \cite{guth81,linde94,linde83} and modifications of general relativity \cite{nojiri2017,nojiri2011,nojiri2006a,capozziello2011,faraoni2010,olmo2011,oikonomou2023,gadbail2024}. While both perspectives are valuable, the modified gravity approach is often regarded as more encompassing, as it naturally incorporates key theoretical features relevant to inflation.
\par
String theory, as mostly comprehensive high energy extension of general relativity and the Standard Model, is expected to leave imprints on the low energy effective description of inflation. Since the inflationary phase is essentially classical, with the universe evolving in four dimensions, it is natural to consider frameworks where string-inspired corrections appear in the inflationary Lagrangian.
\par
In this context, the four-dimensional Einstein–Gauss–Bonnet (EGB) theory \cite{nojiri2006b,cognola2008,nojiri2005a,nojiri2005b,nojiri2024,satoh2008,bamba2014,yi2018,guo2009,guo2010,jiang2013,kanti2015,bruck2018,kanti1999,pozdeeva2020,laurentis2015,chervon2019,fomin2022,nozari2017,khan2022,odintsov2018,venikoudis2021,kawai1998,figueroa2023,yi2019,bruck2016,kleihaus2020,bakopoulos2020a,maeda2012,bakopoulos2020b,ai2020,odintsov2020a,odintsov2020b,odintsov2020c,odintsov2020d,oikonomou2024,pozdeeva2025,easther1996,antoniadis1994,chakraborty2024,antoniadis1991,kanti1996,mavromatos2023,kanti1998} plays a prominent role. This framework extends canonical scalar field inflation by including Gauss–Bonnet terms, which arise as low-energy corrections from string theory. When combined with a scalar potential, the EGB model not only provides a viable mechanism for inflation and early-universe dynamics but also ensures perturbative stability.
\par
From an observational standpoint, inflationary studies are still primarily constrained by two key quantities: the scalar spectral index $n_s$ and the tensor to scalar ratio $r$. Despite progress, many inflationary models remain consistent with current Planck data, leading to significant degeneracies. To distinguish among these models, additional probes such as primordial non-Gaussianity \cite{choudhury2024,papanikolaou2024,yoo2016,cai2022,odintsov2021} are essential. This paper aims to review primordial non-Gaussianity in detail, with particular emphasis on its theoretical properties, observational signatures, and implications within the 4D Einstein–scalar–Gauss–Bonnet framework.
\par
In the quantum field theory’s sense, the power spectrum of inflationary fluctuations corresponds to the value of vacuum expectation of the two-point correlation function, while the simplest manifestation of non-Gaussianity arises from the vacuum expectation value of the three-point function. Non-Gaussianity originates from higher order interactions, typically evaluated using the ‘in-in’ (the Schwinger-Keldysh) perturbative formalism \cite{choudhury2024,yoneya2025}, where the two-point function serves as the foundational input. Ensuring that such corrections remain subdominant to tree-level contributions is critical; otherwise, the perturbative approach—central to setting initial post-inflationary conditions—would break down.
\par
This work investigates the role of non-Gaussianity as a probe of the Einstein–scalar–Gauss–Bonnet inflationary model. The gravity 4D Einstein– Gauss–Bonnet (4DEGB) theory is of particular interest as it emerges naturally from string-theoretic considerations \cite{fernandes2022,ong2022}, representing a promising candidate for describing the primordial universe. Theoretical motivations for this framework stem from its string theory origins, its relevance to inflationary dynamics and the early universe \cite{fasiello2014}, and its ability to preserve the stability of perturbations \cite{hikmawan2016}.

\section{The 4D Einstein-scalar-Gauss-Bonnet (ESGB) of Gravity}
\label{sec2}

The Lagrangian action in the 4D Einstein–Scalar–Gauss–Bonnet (ESGB) inflationary model \cite{nojiri2006b,cognola2008,nojiri2005a,nojiri2005b,nojiri2024,satoh2008,bamba2014,yi2018,guo2009,guo2010,jiang2013,kanti2015,bruck2018,kanti1999,pozdeeva2020,laurentis2015,chervon2019,fomin2022,nozari2017,khan2022,odintsov2018,venikoudis2021,kawai1998,figueroa2023,yi2019,bruck2016,kleihaus2020,bakopoulos2020a,maeda2012,bakopoulos2020b,ai2020,odintsov2020a,odintsov2020b,odintsov2020c,odintsov2020d,oikonomou2024,pozdeeva2025,easther1996,antoniadis1994,chakraborty2024,antoniadis1991,kanti1996,mavromatos2023,kanti1998} can be fully expressed by considering the combination of Einstein gravity, the scalar field, and the Gauss–Bonnet correction term interacting with the scalar field. The explicit mathematical formulation is given as follows:

\begin{equation}
S = \int d^4 x \ \sqrt{-g} \left[ \frac{M_{Pl}^{2}}{2}R-\frac{1}{2}\partial_{\mu}\phi\partial^{\mu}\phi-V(\phi)+\xi(\phi)\mathcal{G} \right],
\end{equation}

\noindent here, $M_{Pl}=(8\pi G)^{-\frac{1}{2}}$, $g$, $R$, $\phi$, and $V(\phi)$ denote the reduced Planck mass, the metric determinant $g_{\mu\nu}$, the Ricci curvature, the scalar inflaton field, and the inflationary potential, respectively. $\xi(\phi)$ is the Gauss-Bonnet coupling, and $\mathcal{G}$ is the 4D Gauss-Bonnet term, expressed as $\mathcal{G}=R^2-4R_{\mu\nu}R^{\mu\nu}+R_{\mu\nu\rho\sigma}R^{\mu\nu\rho\sigma}$.

In this paper, we also adopt the setup in which the cosmological background corresponds to a flat Friedmann-Lemaitre-Robertson-Walker (FLRW) metric with signature $(-+++)$, whose line element is given by: 

\begin{equation}
ds^2 = -dt^2 + a^2(t)\delta_{ij}dx^{i}dx^{j},
\end{equation}

\noindent here, $a(t)$ represents the scale factor. Accordingly, the Ricci scalar and Gauss-Bonnet term takes the forms $R=12H^2+6\dot{H}$ and $\mathcal{G}=24H^2 ( \dot{H}+H^2 )$ where $H=\frac{\dot{a}}{a}$ is the Hubble parameter, and the "dot $ \left( \dot{\,} \right)$" indices the differential have respected to cosmic time. Under the assumption that the scalar field is spatially homogeneous, the term of kinetic reduces to a much simpler expression, denoted by $-\frac{1}{2}\partial_\mu\phi\partial^{\mu}\phi=-\frac{1}{2}\dot{\phi}^{2}$.

In this study, our analysis is confined to the gravitational part of the action, in line with the formulation given in \cite{easther1996}. Specifically, we adopt the standard Einstein–Gauss–Bonnet action while neglecting matter field contributions, thereby concentrating solely on the pure gravity sector.

It is widely recognized that the Gauss-Bonnet (GB) correction term within the Einstein-Hilbert action is topologically invariant, and thus, it does not influence the gravitational field equations in spacetimes of dimension $D<5$. As a result, in four-dimensional spacetime, the GB term has no impact on the dynamics of gravity. Nevertheless, a novel approach to four-dimensional Einstein-Gauss-Bonnet (4D EGB) gravity by employing a dimensional regularization scheme was introduced by \cite{fernandes2022,ong2022,glavan2020}, where the GB coupling constant is rescaled as $\xi \rightarrow \xi/(D-4)$. This modification enables the GB term to contribute nontrivially in the limiting case $D\rightarrow4$.

\section{Arnowitt-Deser-Misner 3+1 Decomposition}
\label{sec3}

Small perturbation of the homogenous section of the inflationary scalar field (inflaton) $\bar{\phi}$, then the metric in Eq. (2) can be written as,
\begin{eqnarray}
\label{3}
\phi(x,t) = \bar{\phi}(t)+\delta\phi(x,t), \nonumber \\
ds^2=-N^2 dt^2 + \gamma_{ij}(dx^{i}+N^{i}dt)(dx^{j}+N^{j}dt),
\end{eqnarray}
\noindent $\gamma_{ij}$ is metric of three-dimensional with the constant $t$ hypersurfaces, $N$ is the lapse function, and $N^{i}$ is the shift vector.

The Einstein and scalar parts (with the exact Arnowitt-Deser-Misner (ADM) metric) obtained are  
\begin{eqnarray}
\label{4}
E_{EH+\phi}&=&\int{dt \, d^3x} \, N\sqrt{h} \;  \biggl[ \frac{M^{2}_{Pl}}{2} \left( R^{(3)} + K_{ij}K^{ij} - K^2  \right) \nonumber \\ &\,& \qquad \quad + -\frac{1}{2N^2} \left( \dot{\phi}-N^i\partial_i\phi \right)^2 - \frac{1}{2}h^{ij}\partial_i \phi \partial_j \phi -V(\phi) \biggr].
\end{eqnarray}

In four dimensions, $\mathcal{G}$ itself is topological; dynamics arise when it is coupled to a non-constant function $\xi(\phi)$. The full ADM expression for a general $\xi(\phi)\mathcal{G}$ is very lengthy. For cosmological or inflationary applications, it is standard practice to use the unitary gauge (with a homogenous inflaton $\phi=\phi(t)\rightarrow\partial_i\phi=0$), so that $\xi=\xi(t)$. Neglecting the total derivative or boundary terms, the relevant ADM contribution to the Lagrangian can be written compactly as:
\begin{eqnarray}
\label{5}
\mathcal{L}^{ADM}_{GB} &\simeq& N\sqrt{h} \biggr[ -8\dot{\xi}G^{(3)}_{ij}K^{ij}-8\ddot{\xi} \left( K_{ij}K^{ij}-K^{2} \right) \nonumber \\ &\,& \qquad \quad + \frac{4}{3} \dot{\xi} \left( K^{3} - 3KK_{ij}K^{ij} + 2{K_{i}}^{j} {K_{j}}^{k} {K_{k}}^{i} \right) \biggr].
\end{eqnarray}

\section{Scalar perturbation analysis}
\label{sec4}

The work is driven by a \textit{comoving gauge},

\begin{equation}
\label{6}
\delta\phi(x,t)=0, \quad \gamma_{ij}(x,t)=a^2(t)[1+2\zeta(x,t)]\delta_{ij},
\end{equation}

\noindent where, $N^i$ and $N$ represent the solutions to the momentum and Hamiltonian constraint equations respectively. Tensor perturbations are disregarded in this analysis. After performing the necessary algebraic steps, the variable of constant density curvature $\zeta(x,t)$ \cite{durrer2024} at the perturbation second-order action is expressed as
\begin{equation}
\label{7}
S^2=M^{2}_{pl} \int{dt \, d^3x} \, a^3 \frac{\epsilon}{c^{2}_{s}} \left[ \dot{\zeta}^2-\frac{c^{2}_{s}}{a^2}(\partial_i \zeta)^2 \right],
\end{equation}

\noindent the overdot $(\dot{\:})$ indicates the differential using cosmic time, and $\epsilon\equiv-\dot{H}/H^2$ denotes the Hubble parameter of slow-roll inflation, which remains nearly constant throughout inflation. The definition of sound speed $c_s$ is described by,
\begin{equation}
\label{8}
c^{2}_{s}=\frac{P_{,X}}{P_{,X}+2XP_{,XX}},
\end{equation}

\noindent where a comma denotes as partial derivative.

Prior to evaluating the corrections to the power spectrum inflation, we first revisit the normal criterion of cosmological perturbation quantization using the variable of Mukhanov-Sasaki (MS) \cite{mukhanov1981,sasaki1986},

\begin{equation}
v\equiv M_{\text{pl}} z \zeta, \qquad z\equiv\frac{a}{c_s}\sqrt{2\epsilon}
\end{equation}

under which the second-order action of Eq. (7) attains canonical normalization

\begin{equation}
S^2= \frac{1}{2} \int{d\tau \, d^3x} \left[ (v')^2-c^{2}_ {s}(\partial_i v)^2 + \frac{z''}{z}v^2 \right],
\end{equation}

\noindent the time of conformal signed by $\tau$, and the prime $(\acute{\;\;})$ indicates the differentiation of $\tau$. During the inflation era, to leading order in the parameter of slow roll, we have:

\begin{equation}
aH=-\frac{1}{(1-\epsilon)\tau}, \quad a(\tau)\propto \tau^{-(1+e)}, \quad \text{and} \quad \frac{z''}{z}=\frac{2+3\epsilon}{\tau^2} 
\end{equation}

Accordingly, the Mukhanov-Sasaki (MS) variable can be expanded in operators as: $\hat{a}^{\dagger}_{-p}$ and $\hat{a}_{p}$ refers to the operator of creation and annihilation, respectively, which conform to the commutation context,

\begin{equation}
\left[ \hat{a}_p , \hat{a}^{\dagger}_{-q}\right] = (2\pi)^3 \delta^3 (p+q)
\end{equation}

The mode functions are associated with the vacuum of Bunch-Davies $|0\rangle$ at the beginning of times, and have meaning by the annihilating condition by the operator $\hat{a}_{p}$. For notational convenience, we represent the constant density curvature perturbation $\zeta$ in two-point form and the power spectrum $\Delta^{2}_{s}$ at the end stage of the inflation era $\tau_0$ as,

\begin{equation}
\langle \zeta(p) \zeta(q) \rangle = (2\pi)^3 \delta^3 (p+q) \langle\langle \zeta(p) \zeta(-p) \rangle\rangle
\end{equation}

\begin{equation}
\Delta^{2}_{s} \equiv \frac{p^{3}}{2\pi^2}  \langle\langle \zeta(p) \zeta(-p) \rangle\rangle
\end{equation}

\noindent We consider the limit $\tau_{0}$, at which point all relevant modes have entered the era of super-horizon and the observable power spectrum $\Delta^{2}_{s}$ has been determined. The angle brackets represent the vacuum expectation value, $\langle \cdots \rangle = \langle 0 | \cdots | 0 \rangle$, while is $\Delta^{2}_{s}$ weighted by the density of the phase space.

The power spectrum can be expressed as 

\begin{equation}
\Delta^{2}_{s(0)} (p)= \left( \frac{H^2}{8\pi^2M^{2}_{pl}c_{s}\epsilon} \right) = \Delta^{2}_{s(0)} \left( p_{*} \right) \left( \frac{p}{p_{*}} \right)^{n_s-1},
\end{equation}

\noindent and the function of modes can be accurately described as

\begin{equation}
\zeta_p(\tau)=\frac{v_p(\tau)}{zM_{pl}}=\left( \frac{H^2}{4M^2_{pl}\epsilon c_s} \right)^{\frac{1}{2}}_H \frac{e^{-ic_sp\tau}}{p^{3/2}} (1+ic_sp\tau),
\end{equation}

\noindent up to leading order in the slow-roll inflation parameter, the subscripts $H$ and $(0)$ indicate the quantities evaluate at the sound-horizon crossing and the contribution of the tree-level, where $c_s p=aH$. The pivot momentum $p_*$ was selected arbitrarily.

The power spectrum $\Delta_s^2$ is nearly scale-invariant, with deviations captured by the spectral index, $n_s-1=O(\epsilon)$. A mild momentum dependence arises from the time variation of the terms inside the brackets at the moment when each mode exits the horizon.

By extending the analysis beyond the second-order action, one can proceed to the third-order action (interaction order) expansion. In this regime $c_s\ll1$, and using the set $\lambda=(2/3) X^3 P_{,XXX}+X^2 P_{,XX}$, the interaction order takes the form \cite{chen2009,braglia2022,chen2013},

\begin{equation}
S_{\text{int}}= \int{dt \, d^3x} \left[ -\frac{2\lambda}{H^3}a^3\dot{\zeta}^3+\frac{\epsilon M^2_{pl}}{Hc^{2}_{s}}a\dot{\zeta} (\partial_i\zeta)^2 \right],
\end{equation}

\noindent This outcome is likewise anticipated in the inflationary effective field theory \cite{cheung2008,weinberg2008,bastero2021,salcedo2024}, where the two interaction terms are independently treated, ensuring that the observables produced by one term do not influence those arising from the other. The resulting interaction order (third order) action corresponds to a cubic self-interaction of $\zeta$.

\section{Evaluating the terms with \textit{'in-in'} formalism}
\label{sec5}

In this step, all cubic terms in $\zeta$ after expanding to third order perturbation are evaluated using \textit{'in-in'} formalism to obtain the correlation function used to define the final form of bispectrum $B$. The evaluation results are presented in the following equations below. Terms 1, 2, and 3 are expressed by following equations Eq. (18), (19), and (20), respectively.

\begin{eqnarray}
\label{18}
\langle \zeta(\textbf{k}_1,0) \zeta(\textbf{k}_2,0) \zeta(\textbf{k}_3,0) \rangle_1 &=& (2\pi)^3 \delta (\textbf{k}_1+\textbf{k}_2+\textbf{k}_3) \frac{H^4}{16\epsilon} \frac{1}{(k_1k_2k_3)^3} \nonumber \\ &\,& {\left( \frac{(k_2k_3)^2}{K} +\frac{k_1(k_2k_3)^2}{K^2}+\text{sym.}\right) } 
\end{eqnarray}

\begin{eqnarray}
\label{19}
 \langle \zeta(\textbf{k}_1,0) \zeta(\textbf{k}_2,0) \zeta(\textbf{k}_3,0) \rangle_2  &=& (2\pi)^3 \delta(\textbf{k}_1+\textbf{k}_2+\textbf{k}_3) \frac{H^4}{32\epsilon^2} \frac{1}{(k_1k_2k_3)^3} \nonumber \\ &\,& \times \biggl\{ (k^2_1+k^2_2+k^2_3) \Bigl[ -K+\frac{1}{K}(k_1k_2+k_1k_3+k_2k_3) \nonumber \\ &\,& \qquad + \frac{1}{K^2}k_1k_2k_3 \Bigr] \biggr\}.
\end{eqnarray}

\begin{eqnarray}
\label{20}
\langle \zeta(\textbf{k}_1,0) \zeta(\textbf{k}_2,0) \zeta(\textbf{k}_3,0) \rangle_3 &=& -(2\pi)^3 \delta (\textbf{k}_1+\textbf{k}_2+\textbf{k}_3) \frac{H^4}{32\epsilon^2} \frac{1}{(k_1k_2k_3)^3} \nonumber \\ &\,& \times \Biggl\{ \left[ \frac{1}{K} \left( \frac{k^4_2}{k_1}+\frac{k^4_1}{k_2}  \right) + \frac{1}{K^2} \left( \frac{k_3k_2^4}{k_1} + \frac{k_3k^4_1}{k_2} \right) \right] \nonumber \\ &\,& \times  \left[ \frac{1}{K} \left( \frac{k^4_3}{k_1} + \frac{k^4_1}{k_3} \right) + \frac{1}{K^2} \left( \frac{k_2k^4_3}{k_1} + \frac{k_2k^4_1}{k_3}\right) \right] \nonumber \\ &\,& \times \biggl[ \frac{1}{K} \left( \frac{k^4_2}{k_3} + \frac{k^4_3}{k_2} \right) \nonumber \\ &\,& \qquad+ \frac{1}{K^2} \left( \frac{k_1k^4_2}{k_3}  + \frac{k_1k^4_3}{k_2}\right) \biggr] \Biggl\}.
\end{eqnarray}

\noindent The correction term is a permutation from, 

\begin{eqnarray}
\label{21}
\langle \zeta(\textbf{x}_1) \zeta(\textbf{x}_2) \rangle \langle \zeta(\textbf{x}_1) \zeta(\textbf{x}_3) \rangle &=&  \int \frac{d^3k_1}{(2\pi)^3} \frac{d^3k_2}{(2\pi)^3} e^{i\textbf{k}_2\cdot(\textbf{x}_2-\textbf{x}_1)} e^{i\textbf{k}_2\cdot(\textbf{x}_3-\textbf{x}_1)} \nonumber \\  &\,& v_{\textbf{k}_1}(\tau) v^*_{\textbf{k}_1}(\tau) v_{\textbf{k}_2}(\tau) v^*_{\textbf{k}_2}(\tau),
\end{eqnarray}

\noindent sym. $=1\leftrightarrow2$ and $=1\leftrightarrow3$. As before, the mode functions are evaluated in the late-time limit, $\tau\rightarrow0$, where they become constant. Thus, Eq. (21) reduces to:

\begin{eqnarray}
\label{22}
\langle \zeta(\textbf{x}_1) \zeta(\textbf{x}_2) \rangle \langle \zeta(\textbf{x}_1) \zeta(\textbf{x}_3) \rangle &=& \int{ \frac{d^3k_1}{(2\pi)^3} \frac{d^3k_2}{(2\pi)^3} \frac{d^3k_3}{(2\pi)^3} e^{i\textbf{k}_1\cdot(\textbf{x}_1)} e^{i\textbf{k}_2\cdot(\textbf{x}_2)} e^{i\textbf{k}_3\cdot(\textbf{x}_3)} } \nonumber \\ &\,& \times (2\pi)^3 \delta(\textbf{k}_1+\textbf{k}_2+\textbf{k}_3) \frac{H^4}{16\epsilon^2} \frac{1}{(k_1k_2k_3)^3} k^3_3 .
\end{eqnarray}

Two additional terms are concluded by $k_3^3\rightarrow(k_1^3+k_2^3+k_3^3)$. So that, correction term reads,

\begin{eqnarray}
\label{23}
\frac{\eta}{4} \left( \langle \zeta_{\textbf{k}_1} \zeta_{\textbf{k}_2} \rangle \langle \zeta_{\textbf{k}_1} \zeta_{\textbf{k}_3} \rangle + \text{sym.} \right) &=& (2\pi)^3 \delta (\textbf{k}_1+\textbf{k}_2+\textbf{k}_3)  \frac{1}{(k_1k_2k_3)^3} \nonumber \\ &\,& \frac{H^4}{32\epsilon^2} \eta \left( k^3_1+k^3_2+k^3_3 \right).
\end{eqnarray}

At this point, we have evaluated all of terms in Lagrangian interaction action with \textit{‘in-in’} formula. The terms are terms 1, 2, 3 and correction term as formed in Eq. (18), (19), (20) and (23). Determining bispectrum $B$ and the explanation result of shape function non-Gaussianity of this cosmological inflation model is described at the following section, result and discussion.

\section{Results}
\label{sec6}

Finally, by substituting $\mathcal{L}_{\text{int}}$  into the in-in formalism, the three resulting terms together with the correction term can be combined to derive the bispectrum of perturbations in the context of 4D Einstein-scalar-Gauss-Bonnet inflation. While the pre-factors of the four shape functions remain essentially unchanged, the simplification of the k-dependence is a highly nontrivial step. After carrying out this simplification, from Eq. (24) yields bispectrum $B$,

\begin{equation}
\langle \zeta_{\textbf{k}_1}\zeta_{\textbf{k}_2}\zeta_{\textbf{k}_3} \rangle = \langle \zeta(\textbf{k}_1)\zeta(\textbf{k}_2)\zeta(\textbf{k}_3) \rangle + \frac{\eta}{4} \left( \langle \zeta_{\textbf{k}_1} \zeta_{\textbf{k}_2} \rangle \langle \zeta_{\textbf{k}_1} \zeta_{\textbf{k}_3} \rangle + \text{sym.} \right)
\end{equation}

\begin{eqnarray}
\label{25}
B \left( k_1,k_2,k_3 \right) &=& 
\frac{H^4}{32\epsilon^2} \frac{1}{(k_1k_2k_3)^3} \Biggr[ (\eta-\epsilon) \left( k^3_1+k^3_2+k^3_3 \right) 
\nonumber \\ &\,& + \epsilon\biggl( k^2_1 k_2 + k^2_2 k_1  + k^2_1 k_3 + k^2_3 k_1 + k^2_2 k_3 + k^2_3 k_2 \biggr) \nonumber \\ &\,& + \frac{8\epsilon}{K} \left( k^2_1 k^2_2 + k^2_1 k^2_3 + k^2_2 k^2_3 \right) \Biggr] .
\end{eqnarray}

To determine the amplitude of non-Gaussianity produced during this inflationary regime, it must first be normalized with respect to the power spectrum, which removes the $\frac{1}{\epsilon^2}$   factor in the Maldacena shape function $B^{\text{mald}}$. Accordingly, upon evaluating this normalized shape function, the superposition of the local shape function and the equilateral shape function is obtained as follows,

\begin{eqnarray}
\label{26}
S(k_1,k_2,k_3) &\approx& \frac{1}{k_1k_2k_3} \biggl[(\eta-\epsilon) \left( k^3_1+k^3_2+k^3_3 \right) \nonumber \\ &\,& + \epsilon \left( k^2_1k_2 + k^2_2k_1 + k^2_1k_3 + k^2_3k_1 + k^2_2k_3 + k^2_3k_2 \right) \nonumber \\ &\,& + \frac{8\epsilon}{(k_1+k_2+k_3)} \left( k^2_1k^2_2 + k^2_1k^2_3 + k^2_2k^2_3 \right)\biggr],
\end{eqnarray}

\begin{eqnarray}
\label{27}
S(k_1,k_2,k_3) &\approx& 2(3\epsilon-\eta) \left( \frac{k^2_3}{k_1k_2}+\frac{k^2_2}{k_1k_3}+\frac{k^2_1}{k_2k_3} \right) \nonumber \\ &\,& + \frac{5}{3} \epsilon \left( \frac{(k_1+k_2-k_3)(k_1-k_2+k_3)(-k_1+k_2+k_3)}{k_1k_2k_3} \right) \nonumber \\ &=& 2(3\epsilon-\eta) \; S^{\text{local}} + \frac{5}{3} \epsilon \; S^{\text{equil}},
\end{eqnarray}

\begin{equation}
S(k_1,k_2,k_3) \approx \\ (6\epsilon-2\eta) \; S^{\text{local}} (k_1,k_2,k_3) + \frac{5}{3} \epsilon \; S^{\text{equil}} (k_1,k_2,k_3).
\end{equation}

Based on \cite{liguori2010,peron2024,baldi2024}, the dominant shape function in Eq.(28) is $S^\text{local}$, that is 99.7\%. Therefore, the appropriate limit for evaluating this function is the squeezed limit, $k_1\ll k_2\approx k_3$, yielding,

\begin{equation}
\lim_{k_1\rightarrow 0} S(k_1,k_2,k_3) \approx \\ (\eta + 2 \epsilon) \frac{k_2}{k_1}
\end{equation}

\begin{equation}
S \approx \\ \mathcal{O}(\epsilon, \eta) \rightarrow \mathcal{O} (\epsilon, \eta)
\end{equation}

Thus, the conclusion that can be drawn from this work is that the obtained non-Gaussianity is suppressing slow-roll in the squeezed limit. As discussed, Planck 2018 
investigates $f_{\text{NL}}\approx\mathcal{O}(1)$; therefore, the amplitude of the non-Gaussianity considered here does not fall within the experimental bounds but rather within the theoretical limits. 

\section{Compare with other inflation models}
\label{sec7}

The results of this analysis, thus far, indicate that the non-Gaussianity produced by this model is comparable to that obtained from the \textit{toy model}, single scalar field, which has long been used as the standard framework. A comparison between this model and other selected inflationary models is presented in Table 1 below. The comparison is based on the normalized fractional correlations $C[S,S']$.

It can be noted that, up to two significant figures, the local template shows a maximal correlation with the shape function derived from both single-field inflation and the 4D EGB model, as demonstrated in this study. Hence, the Planck constraints on the local template are directly applicable to this model. In contrast, other inflationary scenarios such as DBI and ghost inflation are more effectively described by the equilateral template. Accordingly, based on the Planck data discussed earlier, the 4D Einstein-(scalar)-Gauss-Bonnet inflation model yields a reasonable constraint in which the resulting non-Gaussianity is \textit{slow-roll suppressed in the squeezed limit}—a behavior analogous to that of the single-field scalar inflation model. This finding is supported by the analytical biscpectrum calculations that were carried out to elucidate the non-Gaussian characteristics of the model.

\begin{table}[t]
\centering
\begin{tabular}{l c c c c c c c}
\hline
  \; & Local & Equil. & Flat. & DBI & Ghost & Single Field & 4D EGB\\ \hline 
  Local & 1.00 & 0.46 & 0.62 & 0.5 & 0.37 & 1.00 & 1.00\\
  Equilateral & 0.46 & 1.00 & 0.30 & 0.99 & 0.98 & 0.46 & 0.46 \\  
  Flattened & 0.62 & 0.30 & 1.00 & 0.39 & 0.15 & 0.62 & 0.62\\ \hline
\end{tabular}
\caption{Comparison of normalised fractional correlations, $C[S,S']$, between 4D Einstein-Gauss-Bonnet (4D EGB) model with other inflation models, and selected three primordial shape functions (templates) that \textit{Planck} constrains. Data about other models and selected shape functions obtained based on data Ref. \cite{fergusson2009, carron2024}. }\label{fig1}
\end{table}

\section{Conclusions}
\label{sec8}

The calculation of non-Gaussianities in this inflationary model is carried out using the technique of computing n-point correlation functions with a time-dependent interaction state, namely, the in-in formalism. In particular, the derived in-in expression is evaluated at tree level, which provides the first-order description of any non-linearity. This formula is then employed to compute the generating non-Gaussianities in Einstein-scalar-Gauss-Bonnet inflation. From this calculation, it was found that the bispectrum correlates well with the local template, indicating that non-Gaussianities are produced at superhorizon scales.

This situation is described by the Bogoliubov coefficient, which characterizes the degree of initial excitation. When substituted into the in-in formulation, the resulting non-Bunch-Davis bispectrum exhibits a distinctive signature, namely a peak for flattened triangle configurations in the Fourier space.

In summary, this study finds that generating non-Gaussianity in the 4DEGB model is slow-roll inflation parameter suppression in the squeezed limit. This arises because the dominant bispectrum shape function is of the local type, $S^{\text{local}}$. Given that the non-Gaussianity in this framework exhibits a squeezed configuration, one can adopt the condition $k_1\ll k_2 \approx k_3$. As noted, Planck’s observation in 2018 results constrain $f_{\text{NL}}\approx\mathcal{O}(1)$. Therefore, although the amplitudes obtained in this study fall outside the current experimental sensitivity, they remain consistent with the theoretical expectations. As previously discussed, the reasonable constraint obtained for the 4D Einstein–(scalar)–Gauss–Bonnet inflation model based on this work is that the non-Gaussianity is slow-roll suppressed in the squeezed limit, a constraint also exhibited by the non-Gaussianity of the single-field scalar inflation model.

\vspace{1cm}

\noindent \textbf{Declaration of generative AI and AI-assisted technologies in the writing process}

During the preparation of this work, Afiq Agung used ChatGPT (OpenAI) in order to improve the language and readibility of the manuscript. After using this tool, he reviewed and edited the content as needed and take full responsibility for the content of the publication.

\vspace{1cm}

\noindent \textbf{Acknowledgements}

The first and second author, AA and US, gratefully acknowledge the research grant of BIMA by Ministry of Higher Education, Science, and Technology of the Republic of Indonesia (Kemdiktisaintek) for financial support.

\appendix
\section{The third order action and Hamiltonian interaction}
\label{app1}

\subsection{Gauge-restricted line element}
\label{subsecA1}
From Eq. \eqref{5} yields an equation of perturbed line element with gauge-restricted,
\begin{equation}
ds^2=-[(1+\Psi)^2-\partial_iB\partial^iB]dt^2+2a(t)^2\partial_iBdtdx^i+a(t)^2e^{2\zeta}\delta_{ij}dx^idx^j.
\end{equation}

As working in comoving gauge in Eq. \eqref{6}, we made subtitutions: $E=0$, $\delta\phi=0$, and $\Phi=-\zeta$. The linearised constraint equations are re-expressed as,
\begin{equation}
\label{A2}
\Delta\zeta-3H\dot{\zeta} = -3H^2\Psi+Ha^2\Delta B,
\end{equation}
\begin{equation}
\label{A3}
\dot{\zeta} = H\Psi.
\end{equation}
From \eqref{A2} and \eqref{A3}, having the solutions of comoving gauge,
\begin{equation}
\label{A4}
\Psi=\frac{\dot{\zeta}}{H},
\end{equation}
\begin{equation}
\label{A5}
B=-\frac{\dot{\zeta}}{a^2H} +\epsilon\nabla^{-2}\dot{\zeta},
\end{equation}
which reduces the lapse function $N$ and shift vector $N_i$,
\begin{equation}
\label{A6}
N=1+\Psi \quad \rightarrow \quad N= 1+\frac{\dot{\zeta}}{H},
\end{equation}
\begin{equation}
\label{A7}
N_i=a^2B_{,i} \quad \rightarrow \quad N_i=-a^2\partial_i \left( \frac{\zeta}{a^2H}-\epsilon \nabla^{-2}\dot{\zeta} \right).
\end{equation}

\subsection{Determination of $R$, $K_{ij}$, $K^{ij}$, and $K$}
\label{subsecA2}
After giving comoving gauge 3-metric in Eq. \eqref{6} in 3-Christoffel symbols, it yields,
\begin{equation}
\label{A8}
^{(3)}\Gamma^{k}_{ij}=\delta^{kl} \left( \delta_{ik}\partial_j \zeta + \delta_{jk}\partial_i \zeta - \delta_{ij}\partial_k \zeta  \right),
\end{equation}
where $R$ is built out of these, and takes the form,
\begin{equation}
\label{A9}
R=-\frac{2}{a^2e^{2\zeta}} \left( 2\nabla^2\zeta + (\nabla\zeta)^2 \right).
\end{equation}

Hence, an expression for $K_{ij}$ is found inclusive of $N$ and $N_i$,
\begin{equation}
\label{A10}
K_{ij}=\frac{1}{N} \left[ a^2e^{2\zeta}\delta_{ij} - \partial_{(i}N_{j)} + (2N_{(i}\partial_{j)}\zeta-\delta_{ij}N_k\partial^k\zeta) \right],
\end{equation}
has inverse,
\begin{equation}
\label{A11}
K^{ij}=\frac{1}{a^4e^{4\zeta}} \delta^{ik}\delta^{jl}K_{kl}.
\end{equation}

$K$ is determined, by contracting \eqref{A11} with the 3-metric,
\begin{equation}
\label{A12}
K= \, ^{(3)}g_{ij}K^{ij}=3(H+\dot{\zeta})-\frac{1}{a^2e^{2\zeta}} \left( \partial^kN_k+N_k\partial^k\zeta \right).
\end{equation}

\subsection{Subtitution into the action}
\label{subsecA3}
$R$, $K_{ij}$, $K^{ij}$, $K$ and $^{(3)}g_{ij}$ can be substituted into the ADM action to yield,
\begin{eqnarray}
\label{A13}
S_G = \frac{1}{2} \int{d^4x} \, Na^3e^{3\zeta} &\Biggr[& -\frac{2}{a^2e^{2\zeta}} \left( 2\nabla^2 \zeta + (\nabla\zeta)^2 \right)  - \biggl( 3(H+\dot{\zeta}) \nonumber\\ &\,& - \frac{1}{a^2e^{2\zeta}} \left( \partial^kN_k+N_k\partial^k\zeta \right) \biggr)^2  + \frac{1}{N^2} \frac{1}{a^4e^{4\zeta}} \delta^{ik} \delta^{jl} \nonumber\\ &\,&\Bigl[ a^2e^{2\zeta}\partial_{ij} - \partial_{(i} N_{j)}+  \left( 2N_{(i}\partial_{j)}\zeta - \delta_{ij} N_k \partial^k \zeta \right) \Bigr] \nonumber\\ &\,&\times \Bigr[ a^2e^{2\zeta}\delta_{kl} - \partial_{(k}N_{l)} \nonumber\\ &\,& \quad + (2N_{(k}\partial_{l)} \zeta - \delta_{kl}N_i\partial^i\zeta ) \Bigr]  \Biggr].
\end{eqnarray}
All terms leading order and cubic order in $\zeta$ was isolated. The third order action reads,
\begin{equation}
\label{A14}
S_G[3] = \int{d^4x} \, a\epsilon^2 \left[ a^2\zeta\dot{\zeta}^2 - \zeta(\partial_i\zeta)^2+2a^2\dot{\zeta}(\partial_i\zeta)\partial_i \left( \nabla^{-2}\dot{\zeta} \right) \right] + 2f(\zeta) \frac{\delta\mathcal{L}_2}{\delta\zeta}
\end{equation}
all terms proportional to the first order equation of motion are collected into $f(\zeta)$. But, via field re-definition,
\begin{equation}
\label{A15}
\zeta=\zeta_n-f(\zeta_n),
\end{equation}
makes $f(\zeta)$ terms removed from the action.

Therefore, the 3-point correlator of $\zeta$ is found to gain the following additional terms,
\begin{equation}
\label{A16}
\langle \zeta_{\textbf{k}_1}\zeta_{\textbf{k}_2}\zeta_{\textbf{k}_3} \rangle = \langle \zeta_n(\textbf{k}_1)\zeta_n(\textbf{k}_2)\zeta_n(\textbf{k}_3) \rangle + \frac{\eta}{4} (\langle \zeta_n(\textbf{k}_1)\zeta_n(\textbf{k}_2) \rangle\langle \zeta_n(\textbf{k}_1)\zeta_n(\textbf{k}_3) \rangle + \text{sym.}),
\end{equation}
\begin{equation}
\label{A17}
f(\zeta) := \frac{\eta}{4}\zeta^2.
\end{equation}

The third order action that will be used to determine the interaction Hamiltonian is,
\begin{equation}
\label{A18}
S_G[3] = \int{d^4x} \, a\epsilon^2 \left[ a^2\zeta\dot{\zeta}^2 - \zeta(\partial_i\zeta)^2+2a^2\dot{\zeta}(\partial_i\zeta)\partial_i \left( \nabla^{-2}\dot{\zeta} \right) \right].
\end{equation}

\subsection{the Hamiltonian interaction}
\label{subsecA4}
Hence, the interaction Hamiltonian is determined as,
\begin{equation}
\label{A19}
\mathcal{H}_{\text{int}}=-\mathcal{L}_{\text{int}}+\mathcal{O}(\zeta^4),
\end{equation}
\begin{equation}
\label{A20}
H_{\text{int}}=-\int d^3x \: \mathcal{L}_{\text{int}} = -\int d^3x \: a\epsilon^2 \left[ a^2\zeta\dot{\zeta}^2 - \zeta(\partial_i\zeta)^2+2a^2\dot{\zeta}(\partial_i\zeta)\partial_i \left( \nabla^{-2}\dot{\zeta} \right) \right].
\end{equation}

\section{Evaluating with the \textit{'in-in'} formula}
The \textit{'in-in'} formula,
\begin{equation}
\label{B1}
\langle Q(0) \rangle = \text{Re} \Biggl[ \langle0| -2iQ^I(t) \int^t_{-\infty(1+i\epsilon)} dt' H^I_{\text{int}} |0\rangle \Biggr].
\end{equation}
As de Sitter space, $a\approx-1/(H\tau)$, the limit $\tau \rightarrow0$ results in all modes existing far outside the horizon, $k\tau \ll 0$. Hence, the observable becomes,
\begin{equation}
\label{B2}
Q(0)=\zeta(\textbf{k}_1,0)\zeta(\textbf{k}_2,0)\zeta(\textbf{k}_3,0),
\end{equation}
with the corresponding quantum correlator,
\begin{eqnarray}
\label{B3}
\langle \zeta(\textbf{k}_1,0)\zeta(\textbf{k}_2,0)\zeta(\textbf{k}_3,0) \rangle &=& \text{Re} \Biggl[ \langle 0| -2i \zeta^I(\textbf{k}_1,0)\zeta^I(\textbf{k}_2,0)\zeta^I(\textbf{k}_3,0) \nonumber\\ &\,&\qquad \int^t_{-\infty(1+i\epsilon)} dt' \int d^3x \; \nonumber\\ &\,&\qquad \mathcal{L}_{\text{int}} (\zeta^I(\textbf{x},\tau')) |0\rangle \Biggr] .
\end{eqnarray}

Converting the fields within the $\mathcal{L}_\text{int}$ to Fourier space,
\begin{equation}
\label{B4}
\zeta(\textbf{x},t) = \int \frac{d^3k}{(2\pi)}^3 \bigl[ a_{\textbf{k}}v_k(\tau)e^{i{\textbf{k}}\cdot \textbf{x}} + a^\dagger_{\textbf{k}}v^*_k(\tau)e^{-i{\textbf{k}}\cdot \textbf{x}}\bigr],
\end{equation}
with following redefinition of the Mukhanov variable has been made, $v_k:=zv_k$. As detailed, $\zeta=zv_k$; where $z=\sqrt{2a^2\epsilon}$. It can be seen $\mathcal{L}_{\text{int}}$ is constructed such only contains terms $\mathcal{O}(\zeta^3)$, becomes 6-point correlator. This 6-point correlator of interaction picture fields can be contracted into products of 2-point correlators via Wick's theorem. Thus, the 2-point correlator in the non-interacting limit is required, and takes form,
\begin{eqnarray}
\label{B5}
 \langle 0| \zeta_{\textbf{k}_1}(\tau_1)\zeta_{\textbf{k}_2}(\tau_2) |0 \rangle &=& v_{k_1}(\tau_1) v^*_{k_1}(\tau_2) \langle 0| [a_{\textbf{k}_1}a^\dagger_{\textbf{k}_2}] |0 \rangle \nonumber \\ 
      &=& v_{k_1}(\tau_1) v^*_{k_1}(\tau_2) (2\pi)^3 \delta (\textbf{k}_1-\textbf{k}_2),
\end{eqnarray}

Therefore, the calculation of \eqref{B3} begins, first by noticing that $\mathcal{L}_{\text{int}}$ contributes the following three terms,
\begin{equation}
\label{B6}
\mathcal{L}_{\text{int}} = a\epsilon^2 \biggl[ \underset{1}{\underbrace{a^2\zeta\dot{\zeta}^2}} + \underset{2}{\underbrace{\zeta(\partial_i \zeta)^2}} + \underset{3}{\underbrace{2a^2\dot{\zeta}(\partial_i\zeta) \partial_i \bigl( \nabla^{-2}\dot{\zeta} \bigr)}} \biggr],
\end{equation}
which evaluated separately.
\label{app2}

\subsection{Term 1}
\label{subsecB1}
Substituting term 1 of $\mathcal{L}_{\text{int}}$ into \eqref{B3},
\begin{eqnarray}
\label{B7}
\langle \zeta(\textbf{k}_1,0)\zeta(\textbf{k}_2,0)\zeta(\textbf{k}_3,0) \rangle_1 &=& \text{Re} \Biggl[ \langle 0| -2i \zeta(\textbf{k}_1,0)\zeta(\textbf{k}_2,0)\zeta(\textbf{k}_3,0) \nonumber\\ &\,&\qquad  \int d^3x \times \int \frac{d^3p_1}{(2\pi)^3} \frac{d^3p_2}{(2\pi)^3} \frac{d^3p_3}{(2\pi)^3} \nonumber\\ &\,&\qquad e^{-i(\textbf{p}_1+\textbf{p}_2+\textbf{p}_3)\cdot x} a^2 \epsilon^2 \nonumber\\ &\,&\qquad \zeta (\textbf{p}_1,\tau')\zeta(\textbf{p}_2,\tau')\zeta(\textbf{p}_3,\tau') |0\rangle \Biggr] ,
\end{eqnarray}
there are three out of a possible fifteen contractions survive, and the contractions which don't survive are due to the creation and annihilation vacuum identities: $\hat{a}_k|0\rangle=0, \langle0|\hat{a}^\dagger_k=0$. The three are, 
\begin{eqnarray}
\label{B8}
 &\,&\langle \zeta(\textbf{k}_1,0)\zeta(\textbf{p}_1,\tau') \rangle  \langle \zeta(\textbf{k}_2,0)\zeta'(\textbf{p}_2,\tau')\rangle \langle \zeta(\textbf{k}_3,0) \zeta'(\textbf{p}_3,\tau') \rangle \nonumber \\ &\,&  \qquad =  (2\pi)^9 \delta(\textbf{k}_1-\textbf{p}_1)\delta(\textbf{k}_2-\textbf{p}_2)\delta(\textbf{k}_3-\textbf{p}_3) \nonumber\\ &\,& \qquad \qquad v_{k_1}(0)v^*_{k_1}(\tau')v_{k_2}(0)v'^*_{p_2}(\tau')v_{k_3}(0)v'^*_{p_3}(\tau') .
\end{eqnarray}
by symmetry, two additional contributions under $1\leftrightarrow2$ and $1\leftrightarrow3$. Eq. \eqref{B7} is simplified by substitution of the identity,
\begin{equation}
\label{B9}
\int d^3x \; e^{-i(\textbf{p}_1+\textbf{p}_2+\textbf{p}_3)\cdot x} = (2\pi)^3 \delta(\textbf{p}_1+\textbf{p}_2+\textbf{p}_3),
\end{equation}
which leaves,
\begin{eqnarray}
\label{B10}
\langle \zeta(\textbf{k}_1,0)\zeta(\textbf{k}_2,0)\zeta(\textbf{k}_3,0) \rangle_1 &=& \text{Re} \Biggl[ -4i  \int^t_{-\infty(1+i\epsilon)} d\tau' \int d^3p_1d^3p_2d^3p_3 \nonumber \\ &\,& \qquad a^2\epsilon^2(2\pi)^3 \delta(\textbf{p}_1+\textbf{p}_2+\textbf{p}_3) \nonumber\\ &\,& \qquad \times v_{k_1}(0)v_{k_2}(0)v_{k_3}(0)  v^*_{p_1}(\tau')v'^*_{p_2}(\tau')v'^*_{p_3}(\tau') \nonumber \\ &\,& \qquad \delta(\textbf{k}_1-\textbf{p}_1)\delta(\textbf{k}_2-\textbf{p}_2)\delta(\textbf{k}_3-\textbf{p}_3) \nonumber \\ &\,& \qquad + \text{sym.} \Biggr] ,
\end{eqnarray}
'sym.' denotes the addintional two forms from $1\leftrightarrow2$ and $1\leftrightarrow3$. The mode functions take form,
\begin{equation}
\label{B11}
v_k(\tau)=\frac{H}{\sqrt{4\epsilon k^3}}e^{-ik\tau}(1+ik\tau),
\end{equation}
\begin{equation}
\label{B12}
\lim_{\tau\rightarrow0} v_k(\tau) := v_k(0) = \frac{H}{\sqrt{4\epsilon k^3}}(1+ik\tau),
\end{equation}
\begin{equation}
\label{B13}
v'_k (\tau) = \frac{H}{\sqrt{4\epsilon k^3}}k^2\tau e^{-ik\tau}.
\end{equation}
Substituting into \eqref{B10}, and integrating out the delta functions, yields,
\begin{eqnarray}
\label{B14}
\langle \zeta(\textbf{k}_1,0)\zeta(\textbf{k}_2,0)\zeta(\textbf{k}_3,0) \rangle_1 &=& \text{Re} \Biggl[ -4i \frac{H^6}{(4\epsilon)^3} \frac{1}{(k_1k_2k_3)^3} \frac{\epsilon^2}{(H\tau')^2} \nonumber \\ &\,& \qquad \int^t_{-\infty(1+i\epsilon)} d\tau'  \times (2\pi)^3 \delta(\textbf{k}_1+\textbf{k}_2+\textbf{k}_3)e^{iK\tau'}\nonumber\\ &\,& \qquad (k_2k_3)^2 \tau'^2 (1+ik_1\tau') + \text{sym.} \Biggr] ,
\end{eqnarray}
where $K=k_1+k_2+k_3$. The substitution of $a^2$ for the de Sitter approximation cancels a factor of $\tau'^2$. Then, there will be no contribution from the lower bound of the integral due to the $-\infty(1+i\epsilon)$ in the exponent. Thus, the 3-point correlator for the first term in $\mathcal{L}_{\text{int}}$ becomes,
\begin{eqnarray}
\langle \zeta(\textbf{k}_1,0) \zeta(\textbf{k}_2,0) \zeta(\textbf{k}_3,0) \rangle_1 &=& (2\pi)^3 \delta (\textbf{k}_1+\textbf{k}_2+\textbf{k}_3) \frac{H^4}{16\epsilon} \frac{1}{(k_1k_2k_3)^3} \nonumber \\ &\,& {\left( \frac{(k_2k_3)^2}{K} +\frac{k_1(k_2k_3)^2}{K^2}+\text{sym.}\right) } 
\end{eqnarray}

\subsection{Term 2}
\label{subsecB2}
The second term, $a\epsilon^2\zeta(\partial_i\zeta)^2$, was correlated in a similar way to the first. Considering only $\zeta_{k_1}(\partial\zeta_{k_2})\cdot(\partial\zeta_{k_3})$, where two additional terms were carried through, by symmetry, under $1\leftrightarrow2$ and $1\leftrightarrow3$. One noticed that $(\partial_i\zeta)^2\rightarrow(k_2\cdot k_3)\zeta_{k_2}\zeta_{k_3}$ under a Fourier transform. This pre-factor to $\zeta$ expressed as a magnitude of $k$ by noting that,
\begin{equation}
\label{B16}
(\textbf{k}_1+\textbf{k}_2+\textbf{k}_3)^2=0=k^2_1+k^2_2+k^2_3 + 2\textbf{k}_1 \cdot \textbf{k}_2 + 2\textbf{k}_2\cdot \textbf{k}_3 + 2\textbf{k}_1\cdot \textbf{k}_3,
\end{equation}
by the triangle condition, hence,
\begin{equation}
\label{B17}
\textbf{k}_1\cdot \textbf{k}_2 + \textbf{k}_2\cdot \textbf{k}_3 + \textbf{k}_1\cdot \textbf{k}_3 = -\frac{1}{2}(k^2_1+k^2_2+k^2_3).
\end{equation}

Substituting $\textbf{k}_2\cdot \textbf{k}_3 = -\frac{1}{2}k^2_1$ made in the subsequent work. Following the procedures above, only one term survives the Wick contraction,
\begin{eqnarray}
\label{B18}
\langle \zeta(\textbf{k}_1,0) \zeta(\textbf{k}_2,0) \zeta(\textbf{k}_3,0) \rangle_2 &=& \text{Re} \Biggl[ -4iv_{k_1}(0)v_{k_2}(0)v_{k_3}(0) \epsilon^2(2\pi)^3 \nonumber \\ &\,& \qquad \delta(\textbf{k}_1+\textbf{k}_2+\textbf{k}_3)  \times \int^\tau_{-\infty(1+i\epsilon)} d\tau' a^2 \nonumber \\ &\,& \qquad v^*_{k_1}(\tau')v^*_{k_2}(\tau')v^*_{k_3}(\tau')\Biggr].
\end{eqnarray}

The pre-factor to the time integral, remains the same as $\langle \zeta\zeta\zeta \rangle_1$; the conformal time integral was expected as,
\begin{eqnarray}
\label{B19}
\int^\tau_{-\infty(1+i\epsilon)}d\tau' a^2v^*_{k_1}(\tau')v^*_{k_2}(\tau')v^*_{k_3}(\tau') &=& -\int^\tau_{-\infty(1+i\epsilon)}d\tau' \frac{k^2_1}{2(H\tau')^2} e^{iK\tau} (1+ik_1\tau) \nonumber \\ &\,& (1+ik_2\tau)(1+ik_3\tau).
\end{eqnarray}

Evaluating the integral, taking the limit $\tau\rightarrow0$, and adding the symmetric terms, yields,
\begin{eqnarray}
\label{B20}
&\,& -\int^\tau_{-\infty(1+i\epsilon)}d\tau' \frac{k^2_1}{2(H\tau')^2} e^{iK\tau} (1+ik_1\tau) (1+ik_2\tau)(1+ik_3\tau) \nonumber \\ &\,& \qquad \qquad = -\frac{i}{2H^2}(k^2_1+k^2_2+k^2_3) \nonumber \\ &\,& \qquad \qquad \qquad \Bigl[ -K+\frac{1}{K}(k_1k_2+k_1k_3+k_2k_3) + \frac{1}{K^2}k_1k_2k_3 \Bigr].
\end{eqnarray}

The correlators for the second term of $\mathcal{L}_{\text{int}}$ is,
\begin{eqnarray}
 \langle \zeta(\textbf{k}_1,0) \zeta(\textbf{k}_2,0) \zeta(\textbf{k}_3,0) \rangle_2  &=& (2\pi)^3 \delta(\textbf{k}_1+\textbf{k}_2+\textbf{k}_3) \frac{H^4}{32\epsilon^2} \frac{1}{(k_1k_2k_3)^3} \nonumber \\ &\,& \times \biggl\{ (k^2_1+k^2_2+k^2_3) \Bigl[ -K+\frac{1}{K}(k_1k_2+k_1k_3+k_2k_3) \nonumber \\ &\,& \qquad + \frac{1}{K^2}k_1k_2k_3 \Bigr] \biggr\}.
\end{eqnarray}

\subsection{Term 3}
\label{subsecB3}
Evaluation of $\langle \zeta
\zeta \zeta \rangle_3$ began by converting the final term, $2a^3\epsilon^2\dot{\zeta}(\partial_i\zeta)\partial_i(\nabla^{-2}\dot{\zeta})$, to Fourier space. Because there are three separate differential operators here, $3!=6$ arrangements of the modes will survive the Wick contraction via symmetry (rather than three in the previous two evaluations). The general formula for this conversion to Fourier space becomes,
\begin{equation}
\label{B22}
2a^3\epsilon^2\dot{\zeta}(\partial_i\zeta)\partial_i(\nabla^{-2}\dot{\zeta}) \rightarrow \dot{\zeta}_\textbf{k} \frac{\textbf{k}'\cdot \textbf{k}''}{k'^2} \dot{\zeta}_{\textbf{k}'}\zeta_{\textbf{k}''}.
\end{equation}
The appropriate pre-factor to the time integral due to these six symmetric arrangements becomes,
\begin{eqnarray}
\label{B23}
\frac{\textbf{k}_1\cdot \textbf{k}_2}{k^2_1} +\frac{\textbf{k}_1\cdot \textbf{k}_3}{k^2_1} +\frac{\textbf{k}_2\cdot \textbf{k}_3}{k^2_2} +\frac{\textbf{k}_2\cdot \textbf{k}_1}{k^2_2} +\frac{\textbf{k}_3\cdot \textbf{k}_1}{k^2_3} +\frac{\textbf{k}_3\cdot \textbf{k}_2}{k^2_3} \nonumber \\ = -\frac{1}{2} \Biggl( \frac{k^2_3}{k^2_1} + \frac{k^2_3}{k^2_2} + \frac{k^2_2}{k^2_1} + \frac{k^2_2}{k^2_3} + \frac{k^2_1}{k^2_2} + \frac{k^2_1}{k^2_3} \Biggr).
\end{eqnarray}

However,these pre-factors are grouped into three sets of two, due to the indistinguishable arrangements of $\zeta\dot{\zeta}\dot{\zeta}$, leaving the following three time integrals,
\begin{eqnarray}
\label{B24}
 I_1 = -\frac{1}{2} \left[ \frac{k^2_2}{k^2_1}+\frac{k^2_1}{k^2_2} \right] \int^\tau_{-\infty(1+i\epsilon)} d\tau' \, a^2vk_1(\tau')v'_{k_2}(\tau')v'_{k_3}(\tau'), \nonumber \\  I_2 = -\frac{1}{2}  \left[ \frac{k^2_3}{k^2_1} + \frac{k^2_1}{k^2_3} \right] \int^\tau_{-\infty(1+i\epsilon)} d\tau' \, a^2vk_1(\tau')v'_{k_2}(\tau')v'_{k_3}(\tau'), \nonumber \\  I_3 = -\frac{1}{2} \biggl[ \frac{k^2_2}{k^2_3} + \frac{k^2_3}{k^2_2} \biggr] \int^\tau_{-\infty(1+i\epsilon)} d\tau' \, a^2vk_1(\tau')v'_{k_2}(\tau')v'_{k_3}(\tau').
\end{eqnarray}

These time integrals therefore take the form of the ones solved during the evaluation of the first term in $\mathcal{L}_{\text{int}}$. Hence, the momentum dependence is known as $\tau\rightarrow0$, and these integrals reduce to,

\begin{eqnarray}
\label{B25}
 I_1 = -\frac{1}{2} \left[ \frac{1}{K} \left( \frac{k^4_2}{k_1}+\frac{k^4_1}{k_2}  \right) + \frac{1}{K^2} \left( \frac{k_3k_2^4}{k_1} + \frac{k_3k^4_1}{k_2} \right) \right], \nonumber \\  I_2 = -\frac{1}{2}  \left[ \frac{1}{K} \left( \frac{k^4_3}{k_1} + \frac{k^4_1}{k_3} \right) + \frac{1}{K^2} \left( \frac{k_2k^4_3}{k_1} + \frac{k_2k^4_1}{k_3}\right) \right], \nonumber \\  I_3 = -\frac{1}{2} \biggl[ \frac{1}{K} \left( \frac{k^4_2}{k_3} + \frac{k^4_3}{k_2} \right) + \frac{1}{K^2} \left( \frac{k_1k^4_2}{k_3}  + \frac{k_1k^4_3}{k_2}\right) \biggr] .
\end{eqnarray}

Combining these produces the 3-point correlator of the third term in $\mathcal{L}_{\text{int}}$,
\begin{eqnarray}
\langle \zeta(\textbf{k}_1,0) \zeta(\textbf{k}_2,0) \zeta(\textbf{k}_3,0) \rangle_3 &=& -(2\pi)^3 \delta (\textbf{k}_1+\textbf{k}_2+\textbf{k}_3) \frac{H^4}{32\epsilon^2} \frac{1}{(k_1k_2k_3)^3} \nonumber \\ &\,& \times \Biggl\{ \left[ \frac{1}{K} \left( \frac{k^4_2}{k_1}+\frac{k^4_1}{k_2}  \right) + \frac{1}{K^2} \left( \frac{k_3k_2^4}{k_1} + \frac{k_3k^4_1}{k_2} \right) \right] \nonumber \\ &\,& \times  \left[ \frac{1}{K} \left( \frac{k^4_3}{k_1} + \frac{k^4_1}{k_3} \right) + \frac{1}{K^2} \left( \frac{k_2k^4_3}{k_1} + \frac{k_2k^4_1}{k_3}\right) \right] \nonumber \\ &\,& \times \biggl[ \frac{1}{K} \left( \frac{k^4_2}{k_3} + \frac{k^4_3}{k_2} \right) \nonumber \\ &\,& \qquad+ \frac{1}{K^2} \left( \frac{k_1k^4_2}{k_3}  + \frac{k_1k^4_3}{k_2}\right) \biggr] \Biggl\}.
\end{eqnarray}



\pagebreak

\end{document}